\documentclass{elsart}               

\usepackage[dvips]{epsfig}    

\begin{document}

\begin{frontmatter}

\title{Dark Energy: The Observational Challenge}

\thanks[footnoteinfo]{To Appear in
Wide Field Imaging From Space, New Astronomy Reviews, eds. T. McKay, 
A. Fruchter, E. Linder} 


\author[DW]{David H. Weinberg}\ead{dhw@astronomy.ohio-state.edu}   

\address[DW]{Department of Astronomy, Ohio State University, Columbus, OH 43210}


\begin{abstract}                          
Nearly all proposed tests for the nature of dark energy measure some 
combination of four fundamental observables: the Hubble parameter $H(z)$,
the distance-redshift relation $d(z)$, the age-redshift relation $t(z)$,
or the linear growth factor $D_1(z)$.  I discuss the sensitivity of these
observables to the value and redshift history of the equation of state 
parameter $w$, emphasizing where these different observables are and are not
complementary.  Demonstrating time-variability of $w$ is difficult in most
cases because dark energy is dynamically insignificant at high redshift.
Time-variability in which dark energy tracks the matter density at high 
redshift and changes to a cosmological constant at low redshift is 
{\it relatively} easy to detect. However, even a sharp transition
of this sort at $z_c=1$ produces only percent-level differences in $d(z)$ or
$D_1(z)$ over the redshift range $0.4 \leq z \leq 1.8$, relative to the closest
constant-$w$ model.  Estimates of $D_1(z)$ or $H(z)$ at higher redshift,
potentially achievable with the Ly$\alpha$ forest, galaxy redshift surveys, and
the CMB power spectrum, can add substantial leverage on such models, given 
precise distance constraints at $z<2$. The most promising routes to obtaining 
sub-percent precision on dark energy observables are space-based studies of 
Type Ia supernovae, which measure $d(z)$ directly, and of weak gravitational 
lensing, which is sensitive to $d(z)$, $D_1(z)$, and $H(z)$.
\end{abstract}

\end{frontmatter}

\section{Introduction}
The acceleration of the universe should come as a surprise to anyone who
has ever thrown a ball into the air and watched it fall back to earth.
What goes up must come down.  Gravity sucks.  

Repulsive gravity is possible in General Relativity, but it requires
the energy density of the universe to be dominated by 
a substance with an exotic equation of state, having negative pressure 
$p < -\rho c^2/3$.  Alternatively, one can modify GR
itself, so that cosmic acceleration arises even with ``normal'' 
gravitating components like pressureless matter and radiation.
In this article I will adopt the language of ``dark energy'' solutions
in which GR is preserved and a new component drives acceleration,
but at present I think that a modification
of the theory of gravity must be considered almost equally plausible.

Given the theoretical difficulty of explaining cosmic acceleration,
it is worth emphasizing that the observational evidence for it is
impressively robust.  The results of the high-redshift supernova
campaigns published in the late 1990s \cite{riess98,perlmutter99}
provided the first direct evidence for acceleration, but they were
accepted quickly in part because studies of large scale structure and
CMB anisotropies already argued indirectly in favor of a
cosmological constant (e.g., \cite{efstathiou90,krauss95,ostriker95}).
The supernova evidence has become much stronger on its own terms
\cite{riess04}, and measurements of the first acoustic peak
in the CMB power spectrum \cite{debernardis00,hanany00,bennett03} greatly
strengthen the overall case for dark energy by showing that the 
{\it total} energy density of the universe is close to critical.
This rules out the nearly empty open models that might otherwise be
marginally compatible with the supernova data, and in combination
with dynamical evidence for a low density of clustered matter
($\Omega_m \sim 0.2-0.4$), it demonstrates that there must be some
additional entry in the cosmic energy budget.  
Unless this new component has negative pressure, the inferred ages of globular
cluster stars conflict with the most convincing recent estimates of the 
Hubble constant \cite{freedman01}.
We thus have three largely independent lines of argument for dark energy:
the supernova Hubble diagram, the dynamical evidence for low matter density,
and the age of the universe.  In addition, we have the astonishing success
of the inflationary cold dark matter paradigm in explaining high-precision 
measurements of CMB anisotropy, galaxy clustering, the Ly$\alpha$
forest, gravitational lensing, and other phenomena 
(see \cite{seljak04} for an up-to-date discussion),
a success that vanishes if there is no dark energy component.

From a theoretical point of view, there are three different aspects
of the dark energy puzzle.  The first is the ``old'' cosmological constant
problem: a naive application of quantum field theory suggests that the
energy density associated with vacuum zero-point fluctuations should
be of order one Planck mass per cubic Planck length, which exceeds the
observational bound by $\sim 120$ orders of magnitude.  Since the only
natural number that is $\sim 10^{-120}$ is zero, it is usually assumed
that a correct calculation will someday show that the true value of 
the fundamental vacuum energy is exactly zero or 
vanishingly small.  This leaves us with 
the second problem: what {\it is} causing the universe to accelerate?
Finally, there is the coincidence problem: why is the dark energy density 
comparable to
the matter density today, when most models predict that they were very
different in the past and will be very different in the future?

Faced with these conundra, theorists have proposed an impressive variety
of possible solutions.  The most common class introduces
a new energy component, typically a scalar field, and associates dark 
energy with the the field's potential energy or kinetic degrees of freedom.
These solutions do not address the ``old'' cosmological constant problem,
so some other mechanism must set the fundamental vacuum energy to zero.
Alternatively, the dark energy really could be the fundamental vacuum
energy, and it has its remarkably small value as a consequence of
not-yet-understood aspects of quantum gravity, or because it varies 
widely throughout an inflationary ``multiverse'' and anthropic selection 
restricts life to regions where it is small enough to allow structure formation
\cite{weinberg87,efstathiou95,garriga00}, or because gravitational back 
reaction causes it to oscillate in time producing alternating phases of 
acceleration and deceleration \cite{brandenberger04}.
Finally, there is the possibility that GR itself must be modified,
perhaps pointing the way to extra spatial dimensions or testable 
consequences of string theory.  No matter what, the existence of
cosmic acceleration implies a fundamental revision to our understanding of
the cosmic energy contents, or particle physics, or gravity, or all
of the above.  The combination of theoretical importance and 
observational difficulty make the dark energy problem a worthy 
science driver for an ambitious wide field imaging mission in space.

\section{Constraining Dark Energy}

If dark energy is not a simple cosmological constant, then it is generically
expected to have spatial inhomogeneities that affect
CMB anisotropies and large scale structure.  However, 
unless we are lucky (i.e., dark energy properties are just 
what is required to produce large inhomogeneities), these effects are
too weak to detect in the face of cosmic variance.  
If cosmic acceleration is a consequence of modified gravity,
then laboratory measurements 
or tests for intermediate-range forces could also turn up clues.
But again, we would have to be lucky.

The only generic method of studying dark energy is to measure its influence
on cosmic expansion history, described by the Friedmann equation
\begin{equation} \label{eqn:friedmann}
H(z) = H_0 \left[\Omega_m(1+z)^3 + \Omega_k(1+z)^2 +
                 \Omega_\phi{\rho_\phi(z) \over \rho_{\phi,0}}\right]^{1/2} .
\end{equation}
Here the $\Omega_x$ refer to densities at the present day ($z=0$) in units of
the critical density, and $\Omega_k \equiv 1-\Omega_m-\Omega_\phi$.
I have ignored the radiation term $\Omega_r(1+z)^4$, which becomes
important only at high redshifts.
A ``fluid'' with equation of state $p=w\rho c^2$ has
$\rho_{\phi}(z) = \rho_{\phi,0}(1+z)^{3(1+w)}$, so a true cosmological
constant with $\rho_\phi(z)=\rho_{\phi,0}$ has $w=-1$.
The effects of modified gravity solutions that change the Friedmann
equation itself may be well approximated by a dark energy
model with some $\rho_\phi(z)$, but perhaps not in all cases.

There have been many proposed tests of dark energy based on supernovae,
radio galaxies, the virial masses, X-ray properties, or Sunyaev-Zel'dovich
decrements of galaxy clusters, large scale structure traced by 
galaxies, clusters, or the Ly$\alpha$ forest, 
galaxy ages, or various aspects of strong or weak
gravitational lensing (see \cite{kujat02} for an extensive but
not exhaustive list).
Nearly all of these tests depend on some
combination of four fundamental observables: the Hubble parameter $H(z)$,
the distance-redshift relation $d(z)$, the age-redshift relation $t(z)$,
or the linear growth factor of mass fluctuations $D_1(z)$.
The evolution of $H(z)$ is governed by the Friedmann 
equation~(\ref{eqn:friedmann}).
The comoving line-of-sight distance to an object at redshift $z$ is
\begin{equation} \label{eqn:distance}
d(z) = {c \over H_0} \int_0^z dz' {H_0 \over H(z')}~.
\end{equation}
In a flat ($\Omega_k=0$) universe, the angular diameter and luminosity
distances measured with standard rulers or candles are related to 
$d(z)$ by cosmology-independent powers of $(1+z)$.  In a curved universe,
they are proportional instead to the transverse comoving distance,
\begin{equation} \label{eqn:distance2}
d_M(z) = {c \over H_0} |\Omega_k|^{-1/2} 
                       S_k\left(|\Omega_k|^{1/2} {H_0 \over c} d(z)\right)~,
\end{equation}
where $S_k(x)={\rm sin}x$ or ${\rm sinh}x$ for $k=+1$, $-1$ \cite{hogg99}.
The age of the universe at redshift $z$ is
\begin{equation} \label{eqn:age}
t(z) = \int_z^\infty {dz' \over (1+z')H(z')} ~.
\end{equation}
The linear growth factor is the solution to the differential equation
\begin{equation} \label{eqn:lingrowth}
\ddot D_1 + 2H(z) \dot D_1 - {3\over 2}\Omega_m H_0^2 (1+z)^3 D_1 =0 ~.
\end{equation}
The solution can be written as a simple integral only for special
values of $w$ (including $w=-1$), but one can gain intuition using
the approximation $d\log D_1/d\log a \equiv f(\Omega_m) \approx 
\Omega_m^{4/7}$, implying
\begin{equation} \label{eqn:lin_approx}
\log D_1(z) = -\int_0^z {dz \over 1+z} [\Omega_m(z)]^{4/7} ~,
\end{equation}
where 
$\Omega_m(z) = \Omega_m(1+z)^3 {\bigr /}
  \left[ \Omega_m(1+z)^3 + \Omega_k(1+z)^2 + 
         \Omega_\phi\rho_\phi(z)/\rho_{\phi,0}\right]$
is the matter density parameter at redshift $z$.
Here I have normalized $D_1(z=0)\equiv 1$ and used the index $4/7$
that is accurate for $\Omega_m(z)\approx 1$ \cite{lightman90} 
instead of the value 0.6 that is more accurate for low $\Omega_m$
\cite{peebles80}.
Figure~1 shows that the approximation~(\ref{eqn:lin_approx}) 
is accurate to better than 1\% for a wide range of models;
the accuracy degrades to $\sim 3\%$ using $f(\Omega_m)=\Omega_m^{0.6}$.

\begin{figure}
\begin{center}
\epsfig{file=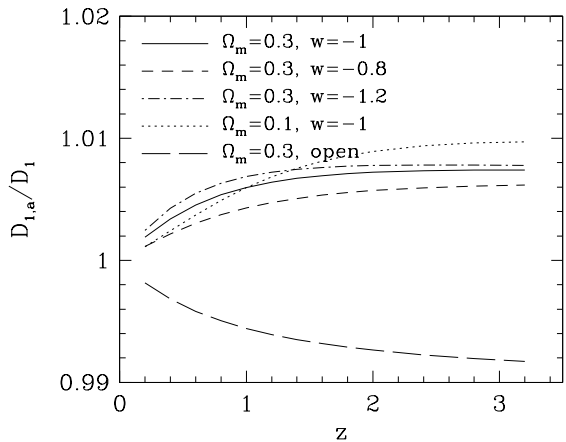,width=3in}
\caption{Ratio of the approximate growth factor $D_1(z)$ calculated by
equation~(\ref{eqn:lin_approx}) to the exact value for various 
combinations of $\Omega_m$ and $w$, including an open universe model
with $\rho_\phi=0$.}
\label{fig:approx}
\end{center}
\end{figure}

The virtue of discussing dark energy tests in terms of these observables
is that one can see whether and how different observational strategies
complement each other.  For example, Type Ia supernovae and other 
standard candle
or standard ruler methods measure $d(z)$ directly, while the abundance of
clusters as a function of redshift depends on both $D_1(z)$ and the 
volume element $dV \propto d^2(z)/H(z)$.  Weak lensing observables
depend on $D_1(z)$, $d(z)$, and $H(z)$, depending on just what 
properties (power spectrum, skewness, etc.) are measured.
The Alcock-Pacyznski \cite{alcock79} anisotropy test measures
the product $d(z)H(z)$.  

\begin{figure}
\begin{center}
\epsfig{file=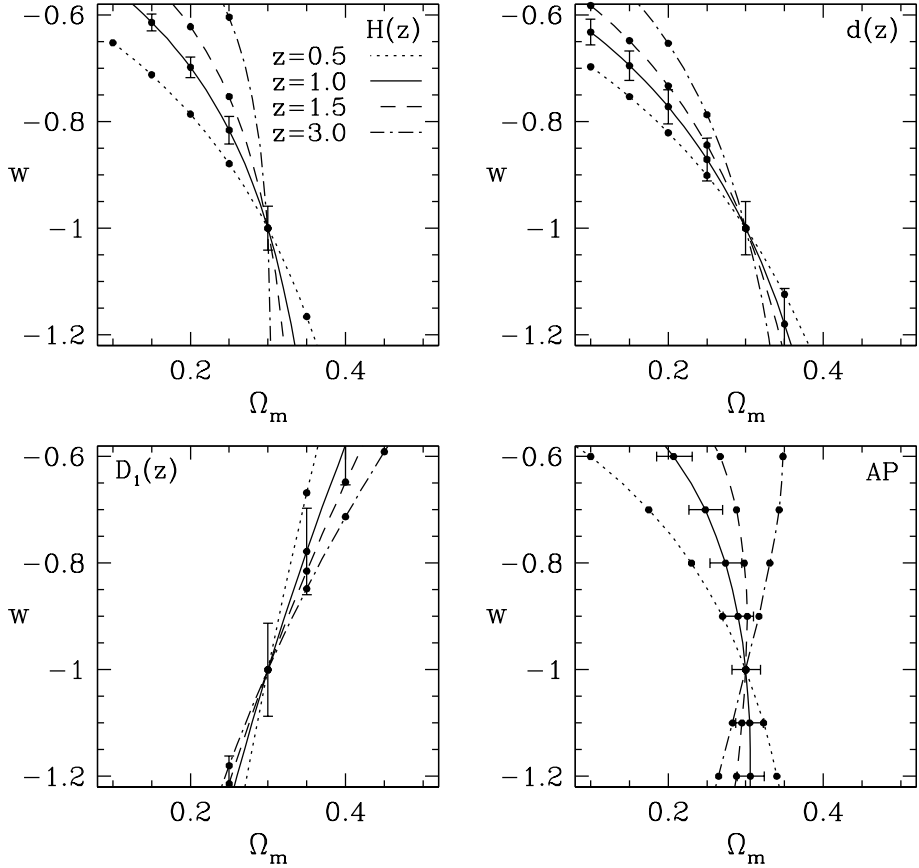,width=5in}
\caption{
Contours in the $\Omega_m-w$ plane along which the values of expansion history
observables are constant, assuming $\Omega_k=0$, constant $w$, and fixed $H_0$.
The four line types correspond to four redshifts, as indicated.
Error bars show the uncertainty in $w$ or $\Omega_m$ associated with a
$\pm 1\%$ uncertainty in the observable, at $z=1$.
}
\label{fig:contours}
\end{center}
\end{figure}

Current observational evidence demonstrates that $\rho_\phi \neq 0$, and it
is compatible with a model having $\Omega_m=0.3$, $\Omega_\phi=0.7$,
$\Omega_k=0$, and $w=-1$.  To show that dark energy is not a cosmological
constant, one must show that the data are incompatible with constant
$\rho_\phi(z)$.  To show that the equation of state parameter
$w$ is time-variable, one must show that the data are incompatible 
with $\rho_\phi(z)\propto (1+z)^n$ for any choice of $n$.  Let us begin with
the first, much less demanding problem, and make the further simplifying
assumption that $\Omega_k=0$.  While CMB measurements provide a precise
value of the angular diameter distance to the surface of last scattering,
they do not demonstrate that $\Omega_k=0$ at the level of precision
of interest here (see, e.g., \cite{tegmark04}), so exact spatial
flatness is a theoretically motivated assumption rather than an
observational constraint.

With these assumptions, the parameters to be constrained are $w$ and
$\Omega_m=1-\Omega_\phi$.  
The decelerating effect of matter is proportional to $\Omega_m$, 
and the accelerating effect of dark energy is proportional to $\Omega_\phi$
but also depends on the pressure-to-density ratio $w$.
For any given redshift and observable, there is a contour in the
$\Omega_m-w$ plane along which the value of the observable is constant.
Figure~2 shows such contours at $z=0.5$, 1, 1.5, and 3 for the Hubble
parameter $H(z)$, the distance $d(z)$, the linear growth factor $D_1(z)$, 
and the Alcock-Pacyznski (AP) parameter $H(z) d(z)$, assuming a fixed value of 
$H_0$.  Error bars on the $z=1$ curves show the impact of a 1\% error
on the corresponding observable.  I have not included plots for
$t(z)$ because I think it unlikely that the systematic errors on absolute age
measurements can be brought low enough to teach us anything we do not
already know about dark energy.

There are several lessons to be drawn from Figure~1.  First, interesting
($\Delta w \sim 0.1$) constraints on $w$ typically require $\sim 1-2\%$ 
measurements at a given redshift, if $\Omega_m$ is known independently.
Second, measurements of different observables, or of the same observable 
at different redshifts, can break the degeneracy between $\Omega_m$
and $w$.  Combinations of $D_1(z)$ with $d(z)$ or $H(z)$ are especially
useful in this regard, because $\Omega_m$ changes have opposite effects.
Applications of the AP test at $z\sim 2-3$, which may be possible with
the Ly$\alpha$ forest \cite{hui99,mcdonald99,mcdonald01}, 
would yield $\Omega_m$ 
constraints that are nearly independent of $w$, also helping to break
degeneracies.  The AP test is independent of uncertainties in $H_0$,
as are measurements of $d(z)$ that use candles or rulers calibrated
in the local Hubble flow and measurements of $H(z)$ that compare
a velocity scale measured at high redshift to a lengthscale measured
from local galaxy redshift surveys.  However, the CMB provides a standard
ruler in absolute Mpc, so tests that use this standard
ruler also require a precise determination of $H_0$ (see \cite{hu04}
for an excellent discussion of this point).  

Constraining $w$ becomes substantially more difficult if we drop the 
assumption that $\Omega_k=0$, since $\Omega_m$ and $\Omega_\phi$ can now vary 
independently.  Most error forecasting papers do not investigate this
case in detail, and I have not done so myself.  Curvature affects all
observables through the $\Omega_k (1+z)^2$ term in the Friedmann equation,
and it additionally affects the distance-redshift relation (and thus
also the AP parameter) through geometrical effects.  Thus, complementary
studies using distance/AP measurements and growth factor/$H(z)$ 
measurements would be especially valuable for this case, as would precise
independent measurements of $\Omega_m$ from large scale structure.
The CMB determines the combination $\Omega_m h^2$ very well, but a tight CMB
constraint on $\Omega_m$ again requires a precise measurement of $H_0$.

Demonstrating time-variation of $w$ is in general very difficult.
If the effective value of $w$ is close to $-1$ today, as the observations
imply, then the ratio of the matter density to the dark energy density
scales as $\sim (1+z)^3$.  Dark energy is therefore dynamically unimportant
even at moderate redshift $(z\sim 2)$, leaving little leverage to show
that the scaling of $\rho_\phi(z)$ is {\it not} adequately described
by a power law $(1+z)^n$ as expected for constant $w$.
While quintessence models generically predict that $w$ varies in time,
the effects of that variation are much too weak to detect in typical
cases \cite{kujat02}.

\begin{figure}
\begin{center}
\epsfig{file=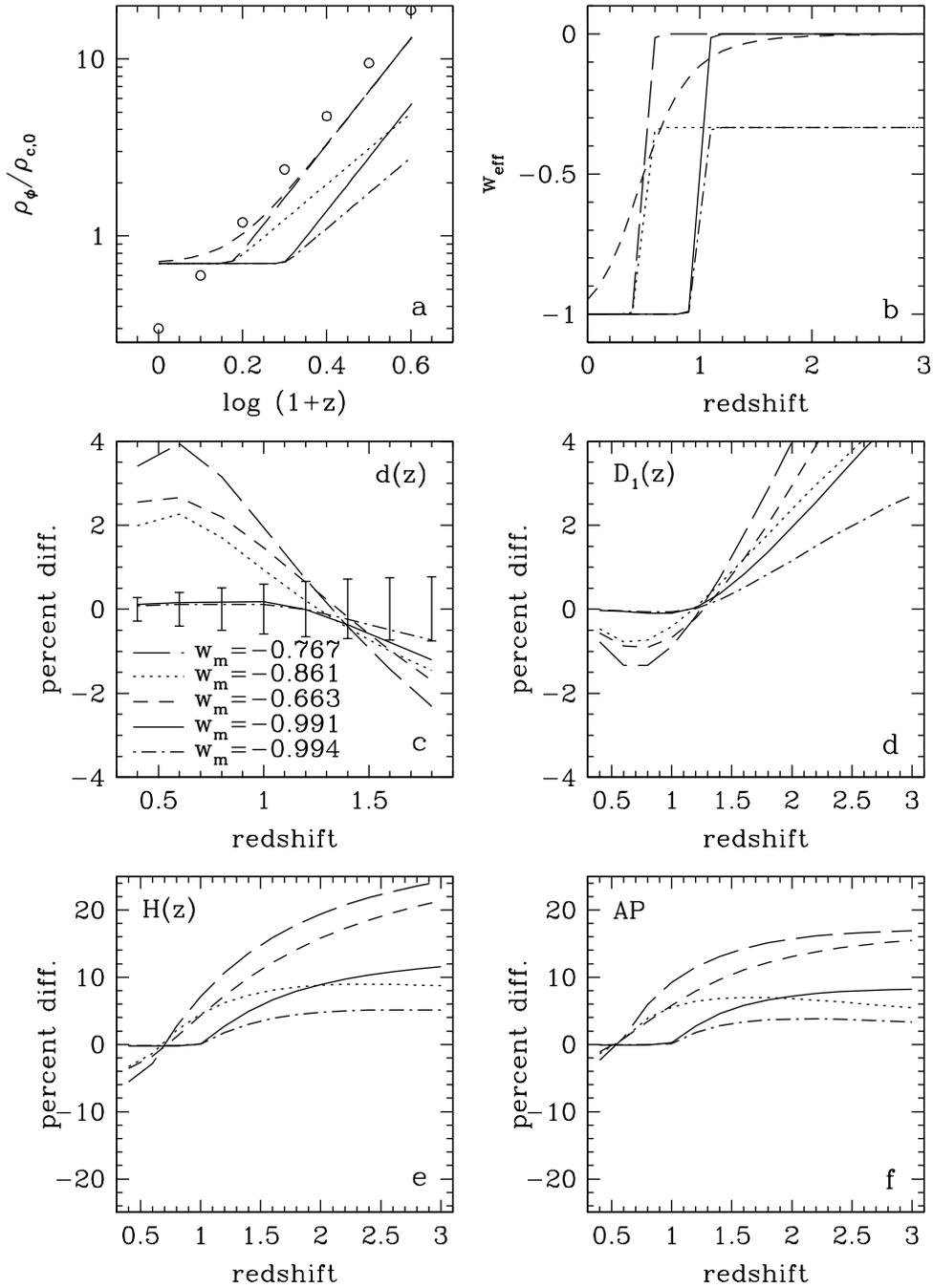,width=5in}
\caption{
Time-varying $w$ models, with the parameterization of 
equation~(\ref{eqn:varyw}).  Panels {\it a} and {\it b} show the evolution
of $\rho_\phi(z)$ and 
$w_{\rm eff} \equiv {1\over 3}d\log\rho_\phi/d\log (1+z) - 1$
for five different parameter combinations.  Circles in panel {\it a} show
the matter density.  Panel {\it c} plots the percentage difference between
the time-variable model and the constant-$w$ model that
best matches its distance predictions over the range $0.4 \leq z \leq 1.8$.
For reference, error bars show the percentage difference caused by changing
$\Omega_m$ by $\pm 0.01$.  Panels {\it d, e, f} show percentage differences in
other observables relative to the same constant-$w$ models.  Note changes in
horizontal and vertical axis scales.
}
\label{fig:varyw}
\end{center}
\end{figure}

Figure~3 illustrates examples from a class of models in which
\begin{equation} \label{eqn:varyw}
\rho_\phi(z) = \rho_{\phi,0} (1+z)^n 
          \left[1+\left({1+z \over 1+z_c}\right)^\alpha\right]^{(m-n)/\alpha} ,
\end{equation}
so that $\rho_\phi \propto (1+z)^n$ at $z \ll z_c$ and 
$\rho_\phi \propto (1+z)^m$ at $z \gg z_c$.
Kujat et al. \cite{kujat02} considered a similar class of models,
but the additional parameter $\alpha$ allows the transition in scaling behavior
at $z_c$ to be arbitrarily sharp, instead of necessarily occurring
over an interval $\Delta z \sim z_c$.
Long-dashed lines show a model that I consider maximally optimistic
from the point of view of detecting time variations of $w$: the energy
scaling changes from $\rho_\phi \propto (1+z)^3$ to $\rho_\phi=\,$constant 
at $z_c=0.5$, and the transition is nearly instantaneous.  One can argue
that $m=3$ is a ``natural'' (or at least interesting) choice because a
model in which $\rho_\phi$ scales like the matter density during the
matter dominated era has a better chance of solving the coincidence problem
\cite{albrecht00,doran01}.  However, the low redshift and sharpness of the 
transition make it as easy as possible to detect, and there is no obvious
reason that nature should be so kind to us.  The short-dashed lines
show a case with a smooth transition at $z_c=0.5$, and solid lines
show a sharp transition at $z_c=1$.  Dotted and dot-dashed lines show
sharp transitions at $z_c=0.5$ and 1, but with high-redshift energy
scaling $\rho_\phi \propto (1+z)^2$ instead of $(1+z)^3$.

Figure~3c shows the key result, focusing on the
redshift range $z=0.4-1.8$ where a space-based Type Ia supernova survey
is likely to produce its tightest constraints.  For each of the
time-variable models, I have identified the constant-$w$ model that
most nearly matches its $d(z)$ relation, minimizing the summed 
absolute fractional differences over the range $0.4 \leq z \leq 1.8$.
The quantity plotted is the percentage difference of $d(z)$ between
the time-variable model and the ``matched'' constant-$w$ model.

For the maximally optimistic model, the prospects look quite good.
There is a 6\% variation in distance between this model and the
matched constant-$w$ model, and a mission like SNAP or DESTINY
could reasonably expect to measure such a variation at high significance.
For a smooth transition (short-dashed line) or high-redshift scaling $m=2$
(dotted line), the variation is $\sim 3\%$, still plausibly within reach.
However, if the transition redshift is $z_c=1$, then the maximum
difference in distance is $\leq 1\%$, and this is reached only in
the highest redshift bin.

Following Figure~2, one might hope that measurements of the growth factor
would complement distance measurements, yielding combined tests of
time variation of $w$ that are much more powerful than either measure
alone.  Unfortunately, the complementarity in the $\Omega_m-w$ space
does not translate to complementarity for this new problem, at least
at $z<2$.  Figure~3d shows percentage differences in $D_1(z)$ relative
to the {\it same} constant-$w$ models that best match the distance
measurements.  
[I have been lazy and used the approximation~(\ref{eqn:lin_approx}).]
The differences in $D_1(z)$ are smaller than those
in $d(z)$, even though the values of $w$ were chosen to match the
latter not the former.  The discriminatory power of $D_1(z)$ does increase
at higher redshifts, especially for $m=3$, because in this case
$\rho_\phi$ freezes in at a constant fraction of $\rho_m$ (Fig.~3a),
so $d{\rm log}D_1/d{\rm log}(1+z) \approx \Omega_m^{4/7}$ is always
depressed below unity.  The Ly$\alpha$ forest might eventually allow
a measurement of $D_1$ at $z\sim 3$ with enough precision to be
interesting in this regard \cite{croft98,mcdonald04}.  The comparison of
CMB anisotropy amplitude to the matter power spectrum amplitude at
low redshift would be more powerful still, if the latter can be
measured precisely enough and the effects of dark energy variation
can be separated from those of the reionization optical depth
and tensor fluctuations, both of which affect the relative
amplitudes of CMB and matter power spectra.

The high-redshift behavior of $\rho_\phi(z)$ has a stronger impact on
$H(z)$ because there is no integrated contribution from low redshift
like those 
in equations~(\ref{eqn:distance}) and~(\ref{eqn:lin_approx}).   Thus
observations that directly constrain $H(z)$ at $z>z_c$ can add 
discriminatory power even if they are much less precise than the distance
measurements at $z<2$.  Because the $H(z)$ variations are much larger
than the $d(z)$ variations, the AP test effectively serves
the same role as an $H(z)$ measurement at high redshift.
The Ly$\alpha$ forest and redshift surveys of Lyman-break galaxies
both offer possible routes to inferring these quantities at the
few percent level 
\cite{weinberg97,hui99,mcdonald99,mcdonald01,blake03,seo03,weinberg03},
though it is not yet clear that systematic errors
in these approaches can indeed be controlled to this level of accuracy.

The experiment in Figure~3 is similar to that carried out by \cite{kujat02}.
If my conclusions here seem a bit more optimistic, it is mainly because
I have zeroed in on the subset of models for which time variation of $w$
is easiest to detect, and partly because the prospect of a wide-field
imager in space makes the idea of percent-level distance measurements
seem worthy of serious discussion, not simply pie in the sky.
Note also that Figure~3 is restricted to flat models.  Evidence for
strong time-variation of $w$ would be surprising enough that one would
want to consider observationally allowed values of $\Omega_k\neq 0$,
thus giving more freedom to reproduce the data with a constant-$w$ model.

\section{Considerations for a Space-Based Dark Energy Experiment}

The discovery of dark energy has profound implications for cosmology
and particle physics, even if we do not yet understand what all of 
those implications are.  A demonstration that the dark energy density
has changed over time (i.e., that $w\neq -1$) would be an achievement of 
comparable magnitude.  A precise measurement of $w$ would rule out many
models of dark energy and provide a clear target for predictive models
that might emerge from fundamental physics.  A demonstration that the
effective value of $w$ has changed with time would be much more remarkable
still, eliminating most models and providing a strong clue to the 
physics of dark energy.

While $w=-1$ has history on its side, it has no 
particular claim to physical plausibility in the absence of a successful
theory of fundamental vacuum energy.  I thus see no reason to conclude
that $w$ is likely to be $-1$ just because current data are consistent
with this value, and improved observations that restricted it to, say,
$w=-1.0 \pm 0.1$ would not change this situation.
If $|w-1|$ is $\sim 0.15$ or more, then there are good prospects for
detecting this departure with the few percent level of precision
that can be obtained by ground-based experiments, especially as different
approaches can provide complementary constraints and independent tests
(Fig.~2).  However, a demonstration that $w=-0.95$ would be just as 
profound as a demonstration that $w=-0.85$, and a demonstration that
$w=-1.05$ would be still more striking, implying physics even weirder
than that we are currently forced to accept.  The level of precision
and control of systematics required to detect such small departures
from $w=-1$ can only be achieved from space.

Detecting time-variation of $w$ would almost certainly require a space-based
investigation, since the complementarity of different methods is largely
lost, and the measurements must probe the largest possible redshift range
to achieve maximum leverage.  Even with the capabilities of a mission
like SNAP or DESTINY, 
I think one must deem a convincing detection of time-variation
unlikely, because for slow variations there will always be a constant-$w$ 
model that produces nearly identical results over the redshift range
where dark energy is dynamically important.  There is one class of 
``generic'' models in which time variation is {\it relatively} easy 
to detect: dark energy scales like matter during the matter-dominated
era and transitions to a constant density when $\Omega_m$ falls 
significantly below one.  However, even in this class the transition
must occur quickly and at low redshift to be observable.

Because I consider a detectable value of $|w-1|$ to be {\it a priori}
much more plausible than a detectable level of time-variation, I think
that one should not sacrifice precision on the former to gain leverage
on the latter.  If one assumes that supernova distance measurements
in redshift bins at $z \leq 1$ will be limited by systematic uncertainty
rather than by statistics, then there is no reason not to pursue higher
redshift objects in the quest for time variability.  This is the
attitude of the SNAP team, and it seems to me entirely reasonable.
However, if cost constraints force consideration of a descoped mission,
then the choice between precision on $w$ and leverage on time-variation
may become unavoidable.  

The very fact that a space-based dark energy mission will be much more
capable than any ground-based experiment raises the stakes, since any 
discovery made by this mission would be unrepeatable for many years to come.
Everyone will have slightly different views on what statistical confidence
level is needed for an ``interesting'' result on dark energy.
In my view, satisfying evidence for $w\neq -1$ would require one 
result with $>3\sigma$ significance or two consistent results with
$>2\sigma$ significance, along with convincing tests for systematic effects.
Time-variation of $w$ demands a higher standard because of its
lower prior probability, maybe one $>4\sigma$ result or two $>3\sigma$.

To my mind, the great advantage of the SNAP concept for a dark energy 
mission relative to the DESTINY concept 
is that the Type Ia supernova experiment and the weak 
lensing experiment look capable of achieving similar statistical precision,
with largely independent systematic uncertainties.
With DESTINY, there is no obvious backup method that comes close to the
precision of the supernova search, making it harder to show that 
an unexpected result on dark energy is actually correct.  
The comparison is not entirely
fair, since DESTINY would likely be much less expensive than SNAP,
and it is no surprise that it has narrower scope.  Either of these
missions would be a great step forward in our efforts to unravel the
most intriguing cosmic mystery of our time, and, as we heard throughout
this workshop, either would enable a vast array of other fascinating
investigations.

\begin{ack}                               
I am indebted to my collaborators on this subject,
Jens Kujat, Angela Linn, and Robert Scherrer.  I also thank the
organizers for inviting me to discuss this topic and for 
patience in awaiting my proceedings contribution.
\end{ack}


\begin{thebibliography}{99}     


\bibitem{riess98}
Riess, A. G., et al.\ 1998, AJ, 116, 1009

\bibitem{perlmutter99}
Perlmutter, S., et al. 1999, ApJ, 517, 565

\bibitem{efstathiou90}
Efstathiou, G., Sutherland, W.~J., \& Maddox, S.~J.\ 1990, Nature, 348, 705 

\bibitem{krauss95}
Krauss, L. M., \& Turner, M. S.\ 1995, Gen Rel \& Grav, 27, 1137

\bibitem{ostriker95}
Ostriker, J.~P.~\& Steinhardt, P.~J.\ 1995, Nature, 377, 600 

\bibitem{riess04}
Riess, A.~G., et al.\ 2004, ApJ, 607, 665

\bibitem{debernardis00}
de Bernardis, P., et al.\ 2000, Nature, 404, 955

\bibitem{hanany00}
Hanany, S., et al.\ 2000, ApJ, 545, L5

\bibitem{bennett03}
Bennett, C.~L.~et al.\ 2003, ApJ Supp, 148, 1 

\bibitem{freedman01}
Freedman, W.~L.~et al.\ 2001, ApJ, 553, 47

\bibitem{seljak04}
Seljak, U., et al.\ 2004, astro-ph/0407372 

\bibitem{weinberg87}
Weinberg, S.\ 1987, Phys Rev Lett, 59, 2607

\bibitem{efstathiou95}
Efstathiou, G.\ 1995, MNRAS, 274, L73

\bibitem{garriga00}
Garriga, J.~\& Vilenkin, A.\ 2000, Phys Rev D, 61, 083502

\bibitem{brandenberger04}
Brandenberger, R.~\& Mazumdar, A.\ 2004, JCAP, 8, 15 

\bibitem{kujat02}
Kujat, J., Linn, A. M., Scherrer, R. J., \& Weinberg, D. H.\ 2002,
ApJ, 572, 1

\bibitem{hogg99}
Hogg, D.W. 1999, astro-ph/9905116

\bibitem{lightman90}
Lightman, A.~P.~\& Schechter, P.~L.\ 1990, ApJ Supp, 74, 831

\bibitem{peebles80} 
Peebles, P. J. E. 1980, The Large Scale Structure of the
Universe (Princeton: Princeton University Press)

\bibitem{alcock79}
Alcock, C., \& Paczy\'{n}ski, B. 1979, Nature, 281, 358

\bibitem{tegmark04}
Tegmark, M., et al.\ 2004, PRD, 69, 103501

\bibitem{hui99}
Hui, L., Stebbins, A., \& Burles, S. 1999, ApJ, 511, 5

\bibitem{mcdonald99}
McDonald, P. \& Miralda-Escud\'e, J. 1999, ApJ, 518, 24

\bibitem{mcdonald01}
McDonald, P. 2001, ApJ, 585, 383

\bibitem{hu04}
Hu, W.\ 2004, astro-ph/0407158 

\bibitem{albrecht00}
Albrecht, A., \& Skordis, C. 2000, PRL, 84, 2076

\bibitem{doran01}
Doran, M., Schwindt, J.-M., \& Wetterich, C. 2001,
Phys Rev D, 64, 123520

\bibitem{croft98}
Croft, R. A. C., Weinberg, D. H., Katz, N., \& Hernquist, L. 1998,
ApJ, 495, 44

\bibitem{mcdonald04}
McDonald, P., et al.\ 2004, astro-ph/0407377 

\bibitem{weinberg97}
Weinberg, D.H., Hernquist, L., Katz, N., Croft, R.
\& Miralda-Escude, J.  1997, in Proc. of the 13th IAP
Colloquium, Structure and Evolution of the IGM from QSO Absorption
Line Systems, eds. P. Petitjean \& S. Charlot, (Paris: Nouvelles
Fronti\`eres), p. 133, astro-ph/9709303 

\bibitem{blake03}
Blake, C.~\& Glazebrook, K.\ 2003, ApJ, 594, 665 

\bibitem{seo03}
Seo, H.~\& Eisenstein, D.~J.\ 2003, ApJ, 598, 720 

\bibitem{weinberg03}
Weinberg, D.~H., Dav{\' e}, R., Katz, 
N., \& Kollmeier, J.~A.\ 2003, AIP Conf.~Proc.~666: The Emergence of Cosmic 
Structure, 666, 157, astro-ph/0301186 


\end{thebibliography}

\end{document}